\documentclass[10pt]{article}
\usepackage{graphicx}
\usepackage{amssymb}
\usepackage{epstopdf}
\title{The role of cell-cell adhesion in wound healing}
\author{Evgeniy Khain$^{1,2}$, Leonard M. Sander$^{1,2}$, \\ and Casey M. Schneider-Mizell $^{2}$ \\
$^1$ Michigan Center for Theoretical Physics\\
$^2$ Department of Physics, University of Michigan}
\date{June 13, 2006}

\begin{document}
\maketitle
\begin{abstract}
We present a stochastic model which describes fronts of cells
invading a wound. In the model cells can move, proliferate, and
experience cell-cell adhesion. We find several qualitatively
different regimes of front motion and analyze the transitions
between them. Above a critical value of adhesion and for small
proliferation large isolated clusters are formed ahead of the
front. This is mapped onto the well-known ferromagnetic phase
transition in the Ising model. For large adhesion, and larger
proliferation the clusters become connected (at some fixed time).
For adhesion below the critical value the results are similar to
our previous work which neglected adhesion. The results are
compared with  experiments, and possible directions of future work
are proposed.
\end{abstract}

\section{Introduction}
When a wound heals, surrounding cells fill the wounded area by
enhanced motility (i.e. diffusion) and enhanced proliferation (see
Ref. \cite{sherratt02} for a recent review). Most  theoretical
treatments of this process \cite{maini04, murray02, sheardown96,
sherratt90} employ a reaction-diffusion equation for the cell
density, equivalent to the Fisher-Kolmogorov (FK) equation
\cite{fisher37, kolmogorov37}. Another approach was taken in
\cite{bramson86, kerstein86}, where a very simple discrete model
was formulated for the similar problem of flame-front propagation.
Recently, this model was applied to wound healing
\cite{callaghan06}. The model takes into account proliferation and
diffusion, and for small proliferation it reduces to the FK
equation. Biologically reasonable proliferation rates are small
compared to rates of diffusion \cite{proliferation}, so the front
velocity is in a good agreement with experimental findings both
for continuum and discrete models. However, the theoretically
predicted width of the front is much larger than the one measured
experimentally (see, for example, Ref. \cite{walker04a}).

Walker et al. \cite{walker04a, walker04} have proposed that the
answer to the paradox lies in the inclusion of cell-cell adhesion.
They investigated an agent based model and observed qualitatively
different regimes of cell organization for low and high values of
adhesion. In order to investigate this idea further, we consider a
simple discrete model which describes the phenomenon of wound
healing, focusing on the key processes: cell-cell adhesion,
diffusion, proliferation. Simulations of our model show two
qualitatively different regimes depending on the adhesion
strength, $q$, similarly to the results of Walker et al.
\cite{walker04a, walker04}. We found that the transition between
these regimes is very sharp, and related to the phase transition
in the Ising model (see below). Another regime of cell
organization was found, depending on the proliferation rate,
$\alpha$. Here we report preliminary results of our study and
analyze the transitions between the regimes. Finally we discuss
the biological applications of our work and compare the results
with experiments.

\section{Formulation of the model}
We will formulate our model in a way which is reminiscent of the
standard `scratch assay' experiment in wound healing studies
\cite{sherratt90}. Consider a square two-dimensional lattice in a
channel geometry. Each lattice site can be empty or once occupied
by a cell.  We assume the lattice distance to be equal to cell
diameter (of the order of $10 \mu m$), taking into account
hard-core exclusion. Thus, a fully occupied region of the lattice
represents a confluent monolayer, that is, unwounded or healed
tissue.

Initially, we put cells into the left part of the channel. We take
$x$ to measure distance along the channel. In the initial state
all sites with $x<40$ are occupied and the rest empty. Thus $x=40$
is the edge of the wound. For $t>0$ cells diffuse and proliferate
along the channel. The dynamical rules which define the model are
as follows: A cell is picked at random, and one of the four
neighboring sites is also picked at random. If this site is empty,
the cell can proliferate to this site (so a new cell is born
there), or migrate there. The probability for proliferation is
$\alpha$. The probability for migration decreases with the number
of nearest neighbors so that $p_{migr} = (1-\alpha)(1-q)^n$, where
$0 \leq q < 1$ is the adhesion parameter, and $1 \leq n \leq 4$ is
the number of nearest neighbors. The case $q=0$ means no adhesion
and brings us back to the simple model \cite{bramson86,
callaghan06, kerstein86}. For nonzero $q$, it is much harder to a
cell to diffuse if it has many neighbors. After each step  time is
advanced by $1/N$, where $N$ is the current number of cells.

As healing proceeds there is a zone in front of the healed tissue
which is partly filled with cells. We call this the invasive
region. We have done extensive simulations of this model. We
report the results in the following sections.

\section{Structure of the invasive region}

\begin{figure}[ht]
\centerline{\includegraphics[width=3.5in,clip=]{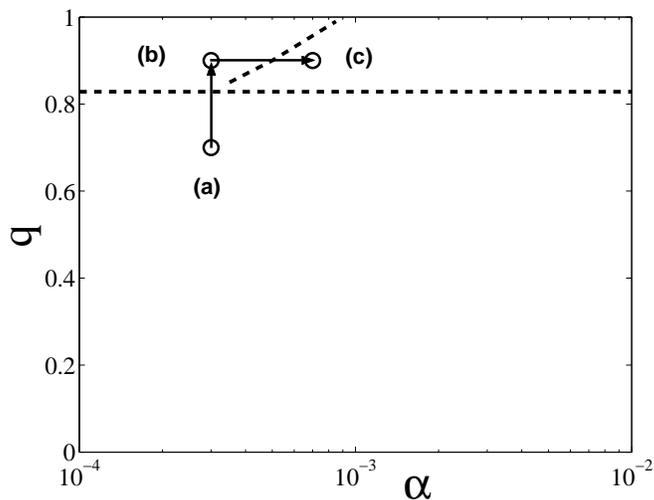}}
\caption{Phase plane $(\alpha, q)$. A qualitatively different
behavior is observed in different regions (a,b,c) in the phase
plane, see Fig.~\ref{snapshots}. Two arrows denote transitions
between (a) and (b), and between (b) and (c), see text.}
\label{phaseplane}
\end{figure}

Our simulations show that the dynamics and  structure of the
invasive zone is qualitatively different depending on the two
parameters of the model: $\alpha$ and $q$. Figure~\ref{phaseplane}
shows the $(\alpha, q)$ phase plane and points out three different
regions of parameters.

\begin{figure}[ht]
\centerline{\includegraphics[width=3.5in,clip=]{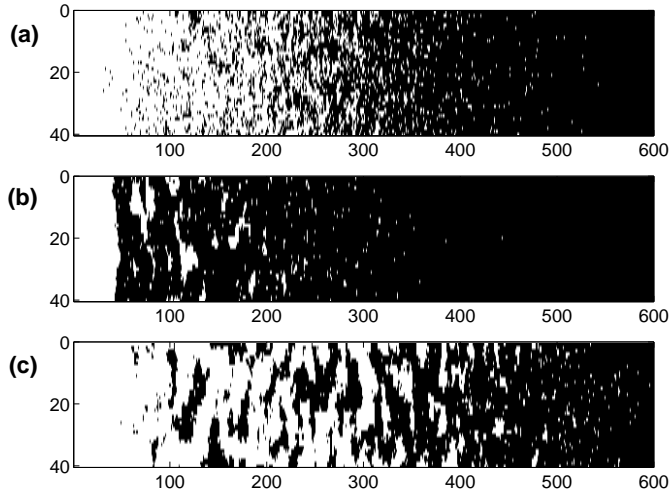}}
\caption{Snapshots of the system in three qualitatively different
regions in the phase plane of parameters, see
Fig.~\ref{phaseplane}. Each white dot corresponds to an occupied
site and each black dot to an empty site; healing proceeds from
left to right. The parameters are $q=0.7, \alpha = 3\times
10^{-4}$ (a), $q=0.9, \alpha = 3\times 10^{-4}$ (b), and $q=0.9,
\alpha = 7\times 10^{-4}$ (c).} \label{snapshots}
\end{figure}

The different types of behavior are shown in
Figure~\ref{snapshots} by means of three snapshots of the system
which correspond to points (a), (b), and (c) in
Figure~\ref{phaseplane}. Figure~\ref{snapshots}a shows the system
for small proliferation and small (subcritical, see below)
adhesion, region (a) in Fig.~\ref{phaseplane}. Here, a front of
cells propagates along the channel, both front velocity and front
width are well-defined (and can be calculated as in
\cite{callaghan06}). Figure~\ref{snapshots}b shows a snapshot for
the same proliferation rate and large (supercritical, see below)
adhesion strength, region (b) in Fig.~\ref{phaseplane}. In
contrast to the propagating fronts shown in Fig.~\ref{snapshots}a,
one can see a number of isolated clusters that are formed in the
invasive region. However, as we increase proliferation (for the
fixed adhesion parameter), another interesting transition occurs.
For moderate proliferation, large clusters in the invasive zone
become connected to each other and to the initial dense front, as
can be seen in Figure~\ref{snapshots}c, region (c) in
Fig.~\ref{phaseplane}.

We can qualitatively analyze the two transitions which are shown
by arrows in Fig.~\ref{phaseplane}. First we focus on the
transition from (a) to (b) which occurs at a fixed (and
sufficiently small, see below) proliferation, when the adhesion
parameter crosses critical value.

We point out that our model \emph{without} proliferation can be
mapped into the Ising model in statistical physics. In this
mapping, an empty site corresponds to spin "down", an occupied
site corresponds to spin "up", so that there is a simple relation
between the average density, $c$, and the average magnetization,
$m$, in the Ising model: $c=(m+1)/2$. The adhesion parameter $q$
is related to the ratio of magnetic coupling $J$ and the
temperature $T$ by $q=1-\exp(-J/k_BT)$. The mapping is possible
because our dynamical rules satisfy detailed balance. Therefore,
the statics of our model is the same as in the Ising model. By
statics, we mean a phase diagram $(m, T)$ (or $(c, q)$ in our
case) which has stable and unstable regions. In the stable region,
a homogenous state (with uniformly distributed cells) remains
homogenous; in contrast, in the unstable region phase separation
occurs and large clusters are formed.

\begin{figure}[ht]
\centerline{\includegraphics[width=3.5in,clip=]{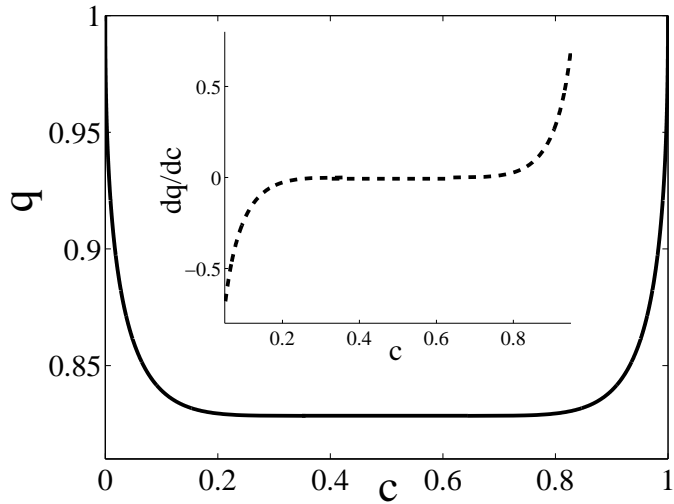}}
\caption{The critical adhesion parameter as a function of density
as given by Eq.~(\ref{Onsager}). An inset shows the derivative
$dq/dc$ versus $c$.} \label{threshold}
\end{figure}

The two-dimensional Ising model was solved  by Onsager
\cite{Onsager}, and the curve $m(T)$, which separates the stable
and unstable regions, is known. In terms of average density $c$
and the critical adhesion parameter $q_c$, we have:
\begin{equation}
c = \frac{1}{2} \pm \frac{1}{2}\left[ 1 -
\frac{16(1-q_{c})^2}{q_{c}^4} \right]^{1/8}. \label{Onsager}
\end{equation}
The unstable region corresponds to $q>q_c$, so for supercritical
adhesion large clusters are formed. Of course, our system is not
homogenous. However the density dependence of $q_c$ is rather slow
in a wide range of intermediate densities, see
Figure~\ref{threshold}. Therefore, we can roughly estimate the
critical adhesion as $q_c\simeq q_c(c=0.5) = 0.8284$. This is the
horizontal dashed line in Figure~\ref{phaseplane}. This
observation explains the different structures observed in the
invasive zones of Figs.~\ref{snapshots}a and \ref{snapshots}b (the
transition from region (a) to region (b),
Figure~\ref{phaseplane}).

\begin{figure}[ht]
\centerline{\includegraphics[width=3.5in,clip=]{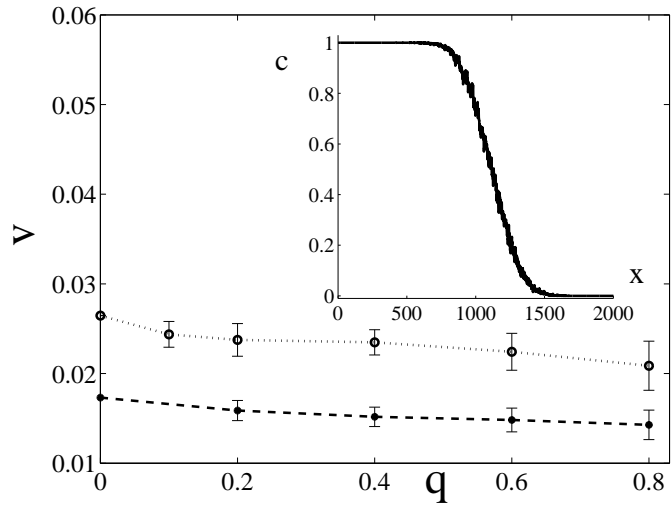}}
\caption{Front velocity as a function of adhesion parameter for
different values of proliferation. The dotted curve corresponds to
$\alpha = 7\times 10^{-4}$, the dashed curve corresponds to
$\alpha = 3\times 10^{-4}$. An inset shows a typical density
profile in region (a). Here $q=0.8$, $\alpha = 3\times 10^{-4}$,
and the time is $t = 10^5$.} \label{velocity}
\end{figure}

The propagation of fronts is a very important topic in statistical
mechanics. It turns out that there are propagating fronts in
region (a), similar to those of the FK equation \cite{fisher37,
kolmogorov37}. We averaged over a series of simulations (and over
the channel width) to obtain smooth density profiles in different
parameter regions. Figure~\ref{velocity} shows an example of such
a front and a velocity of the fronts as a function of adhesion
parameter for different values of proliferation. Note a rather
slow velocity dependence on the adhesion parameter $q$. Clearly,
as $q$ goes to zero, the front velocity $v$ approaches its
theoretical value from the FK equation, which is given in our
notations by $\alpha^{1/2}$. Another important issue is the
scaling properties of the front. For example, it would be
interesting to analyze whether there is a KPZ roughening as in the
case of zero adhesion \cite{KPZ}. Unfortunately, it is extremely
difficult to investigate fronts roughening in this problem,
because the proliferation is very small, so an intrinsic front
width is quite large (see Figure~\ref{velocity}).

\begin{figure}[ht]
\centerline{\includegraphics[width=3.5in,clip=]{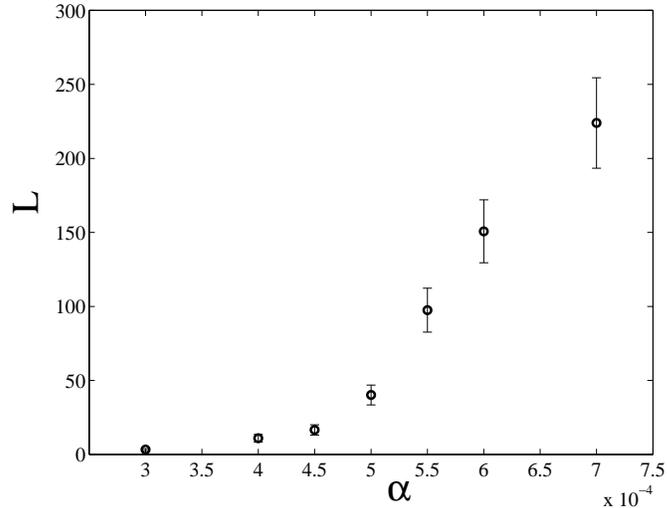}}
\caption{Length of the maximal path along the channel as a
function of proliferation, see text. Note that the proliferation
threshold depends on time: a larger proliferation is needed to get
this percolation-like transition at a smaller time. The adhesion
parameter is $q=0.9$, the time is $t = 3\times10^4$.}
\label{length}
\end{figure}

Now we turn to the transition from region (b) to region (c), see
Fig.~\ref{phaseplane}. We increase the proliferation parameter
$\alpha$, keeping the adhesion parameter $q>q_c$ constant. In
order to compare the results for different values of
proliferation, we fix the total time of the simulations, $t =
3\times 10^4$, and measure the maximum distance $L$ (in $x$
direction) one can move through occupied sites. In other words,
consider a path that passes only through occupied sites. $L$ is
just the $x$ component of the longest path along the channel.
Figure~\ref{length} shows the dependence of $L$ on  $\alpha$. For
small proliferation rate, $L$ is small (as the clusters in the
invasive zone are not connected to each other and to initial dense
region $x<40$) and it slowly increases with $\alpha$ up to a
transition point. But then a small increase in $\alpha$ is
followed by a rapid increase in $L$  as large clusters become
connected, see Figs.~\ref{snapshots}c. Note that the proliferation
threshold depends on time: larger proliferations are needed to get
this percolation-like transition at smaller times. Therefore, the
diagonal dashed line in Fig.~\ref{phaseplane}, which determines
the transition, is not fixed and moves to the right (to larger
proliferations) for smaller times. This indicates that region (b)
is a transient region. For any (biologically reasonable)
proliferation and for supercritical adhesion the system is in
region (b) at early times. However, at very late times there are
propagating fronts with some defined width and velocity. The
fronts develop well inside region (c), much later than the
transition from (b) to (c) occurs. So, the transition from (b) to
(c) is sharp but occurs in a transient regime. The connection to
percolation problem would be an interesting direction of future
research.

\begin{figure}[ht]
\centerline{\includegraphics[width=3.5in,clip=]{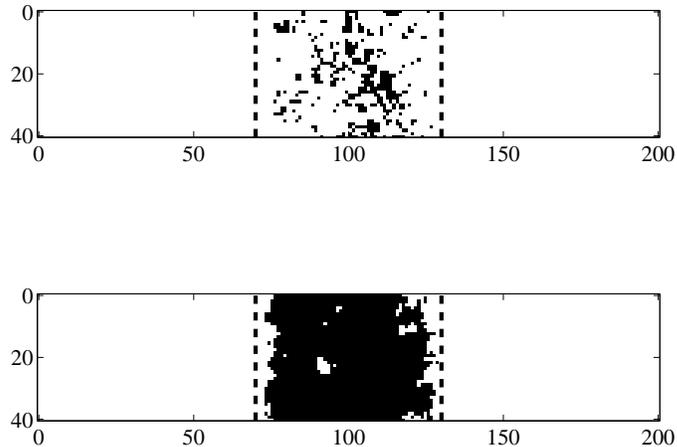}}
\caption{Simulations of wound healing. The upper panel corresponds
to small  adhesion $q=0.7$, the lower panel to
larger adhesion $q=0.99$. White dots represent
occupied sites and  black dots  empty sites. The dashed
lines indicate initial wound edges. The proliferation parameter is
$\alpha=0.01$.} \label{wound}
\end{figure}

To emphasize the effect of cell-cell adhesion, we performed
simulations in the geometry of a scratch assay experiment
\cite{walker04a}. (Up until now we have considered only one side
of the wound.) For the type of cells used by Walker et al.
\cite{walker04a, walker04} adhesion is controlled by the
concentration of Ca$^{++}$ in the system. Lower concentrations of
calcium suppress adhesion, while higher concentrations promote it.

In the experiments \cite{walker04a} an initial width of a scratch
made in a monolayer of epithelial cells was about $600 \mu m$
width, which roughly corresponds to $60$ cell diameters. Since the
characteristic proliferation time (of the order of $1$ day) is
much larger than the characteristic hopping time (of the order of
minutes), the basic time unit in our model approximately equals
the hopping time. Therefore we ran our model out to a  time of
$10^3$, which corresponds to about $24$ hours in the experiment
\cite{walker04a}. Figure~\ref{wound} shows the results of
simulations both for small adhesion strength (upper panel) and
large adhesion strength (lower panel) for the same value of
proliferation rate. For small adhesion  the front is wide and
wound closure is almost complete, whereas for large adhesion there
is a very sharp and slowly moving front, in a good agreement with
the experiments (Figs. 7 and 8 in Ref. \cite{walker04a}) both for
high and low adhesion. There is another point that we would like
to clarify. Figure~\ref{snapshots}c shows clusters of cells
separated by cell-free patches, while Figure~\ref{wound} (lower
panel) does not. This is due to the fact that one should consider
a much larger system (by factor of $6$) and wait for a much longer
time (by factor of $30$) to get the structure shown in Figure
~\ref{snapshots}c.

\section{Summary and discussion}

In this work we presented a stochastic discrete model of wound
healing. In contrast to previous modeling attempts, both discrete
and continuum, we took into account the phenomenon of cell-cell
adhesion. Large adhesion gives rise to sharp  wound healing fronts even
for very low proliferation, see Fig.~\ref{wound}. In
general, depending on the two parameters of the model, the
adhesion $q$ and the proliferation $\alpha$, a completely
different behavior in the invasive zone can be observed.

Certainly, our model oversimplifies a complex biological process
of wound repair. There are many complex biochemical processes such
as calcium waves, apoptosis, etc. which we do not attempt to
treat. One feature which we can treat is the fact that the model
does not take into account a real biological cell cycle which may
influence the dynamics. A cell cycle can be incorporated into the
model in the following way which is based on the work of Walker et
al. \cite{walker04a,walker04}. Each cell is treated as an agent
with its own internal time. A cell moves and adheres according to
the rules described above. Additionally, its internal clock
advances when it is sampled. When a cell reaches the end of the
cell cycle length, mitosis occurs and the cell ceases to move for
a small fraction of the cycle time. It then proliferates in a
random direction with a given probability. In the limit of long
times, the parameter $\alpha$ used in the simple model corresponds
to the product of the division probability and the cycle
frequency. With randomized  initial internal times, this model
behaves very similarly to the purely probabilistic approach which
we have discussed above.

Now we would like to emphasize our main new results, comparing to
Refs.~\cite{walker04a, walker04}. One important new finding,
resulting from our theoretical analysis, is the existence of a
sharp transition between qualitatively different types of behavior
in cell cultures when an adhesion parameter passes a certain
value. Though qualitatively different structures for low and high
adhesion were observed earlier (Refs.~\cite{walker04a, walker04}),
the fact that a very small change in adhesion parameter may give a
qualitatively different behavior was unknown. This result shows a
deep and surprising analogy between ferromagnetic phase transition
in statistical physics and transition to clustering in ensemble of
cells. This analogy had not also been found prior to our work.

Another new result is the transition from region (b) to region
(c), which occurs when increasing proliferation for supercritical
adhesion. This transition and the detailed clusters structure for
supercritical adhesion had not been investigated in
Refs.~\cite{walker04a, walker04}. Finally, there is another new
result, showing a typical front profile of cells density and the
front velocity as a function of adhesion parameter for different
values of proliferation. That possibly gives a way to theoretical
treatment of the problem of cell invasion into a wound in terms of
a modified Cahn-Hilliard equation.

Indeed, a promising direction of future work is continuum modeling
of front propagation, both for subcritical and supercritical
adhesion. A proper candidate here is Cahn-Hilliard equation, which
describes the dynamics of phase separation below the critical
temperature \cite{Godreche, review, Gouyet}. In our system this
corresponds to supercritical adhesion and zero proliferation.
Indeed, one can show \cite{review, Gouyet} that a variant of
Cahn-Hilliard equation can be derived from discrete lattice gas
model with nearest neighbors interaction. This equation applies to
models with conserved order parameter (in our case, the model
without proliferation). One can also try to take into account
proliferation similarly to the FK equation, adding term $\alpha
c(1-c)$ to the Cahn-Hilliard equation. This work is in progress.

\section*{Acknowledgements}
Supported by NSF grant DMS-0244419 and NIH grant CA085139-01A2.


\begin{thebibliography}{10}

\bibitem{bramson86}
M.~Bramson, P.~Calderoni, A.~Demasi, P.~Ferrari, J.~Lebowitz, and
R.~H. Schonmann,
\newblock Microscopic selection principle for a diffusion-reaction equation.
\newblock {\em Journal of Statistical Physics}, {\bf 45}(5-6): 905--920, (1986).


\bibitem{callaghan06}
T.~Callaghan, E.~Khain, L.~M. Sander, and R.~M. Ziff,
\newblock A stochastic model for wound healing.
\newblock {\em Journal of Statistical Physics}, {\bf 122}(5): 909--924, (2006).


\bibitem{fisher37}
R.~A. Fisher,
\newblock The wave of advance of advantageous genes.
\newblock {\em Annual Eugenics}, {\bf 7}: 355–-369, (1937).


\bibitem{Godreche}
J. S. Langer,
\newblock in {\em Solids far from equilibrium}
(ed. C. Godreche) 297--364
\newblock Cambridge, New York, Cambridge
University Press (1992).


\bibitem{review}
J. F. Gouyet, M. Plapp, W. Dieterich, and P. Maass,
\newblock Description of far-from-equilibrium processes by
mean-field lattice gas models.
\newblock {\em Advances in Physics}, {\bf 52}: 523--638, (2003).


\bibitem{kerstein86}
A.~R. Kerstein,
\newblock Computational study of propagating fronts in a lattice-gas model.
\newblock {\em Journal of Statistical Physics}, {\bf 45}(5-6): 921--931, (1986).


\bibitem{kolmogorov37}
A.~Kolmogorov, I.~Petrovsky, and N.~Piscounov,
\newblock Etude de lÕequation de la diffusion avec croissance de la quantite de
  matiere et son application a un probleme biologique.
\newblock {\em Moscow Univ. Bull. Math.}, {\bf 1}: 1--25, (1937).


\bibitem{maini04}
P.~K. Maini, D.~L.~S. McElwain, and D.~Leavesley,
\newblock Travelling waves in a wound healing assay.
\newblock {\em Applied Mathematics Letters}, {\bf 17}(5): 575--580, (2004).


\bibitem{KPZ}
E.~Moro,
\newblock Internal fluctuations effects on Fisher waves.
\newblock {\em Physical Review Letters}, {\bf 87}(23): 238303, (2001).


\bibitem{murray02}
J.~D. Murray,
\newblock {\em Mathematical Biology}.
\newblock Springer, New York (2002).


\bibitem{Onsager}
L.~Onsager,
\newblock Crystal statistics. I. A two-dimensional model with an
order-disorder transition.
\newblock {\em Physical Review}, {\bf 65}: 117--149, (1944).


\bibitem{Gouyet}
M. Plapp and J. F. Gouyet,
\newblock Interface dynamics in a mean-field lattice gas model:
Solute trapping, kinetic coefficient, and interface mobility.
\newblock {\em Physical Review E}, {\bf 55}(5): 5321--5337, (1997).


\bibitem{proliferation} We consider as an example the experiment
of Sheardown and Cheng \cite{sheardown96} on the wounding of
rabbit corneas. It was shown in Ref.~\cite{callaghan06} that the
typical ratio of proliferation rate and basic diffusion rate is of
the order of $3\times10^{-4}$.


\bibitem{sheardown96}
H.~Sheardown and Y.~L. Cheng,
\newblock Mechanisms of corneal epithelial wound healing.
\newblock {\em Chemical Engineering Science}, {\bf 51}(19): 4517--4529, (1996).


\bibitem{sherratt90}
J.~A. Sherratt and J.~D. Murray,
\newblock Models of epidermal wound-healing.
\newblock {\em Proceedings of the Royal Society of London series
B - Biological Sciences}, {\bf 241}(1300): 29--36, (1990).


\bibitem{sherratt02}
J.~A. Sherratt and J.~C. Dallon,
\newblock Theoretical models of wound healing: past successes and future challenges.
\newblock {\em Comptes Rendus Biologies}, {\bf 325}(5): 557--564, (2002).


\bibitem{walker04a}
D.~C. Walker, G.~Hill, S.~M. Wood, R.~H. Smallwood, and
J.~Southgate,
\newblock Agent-based computational modeling of wounded epithelial cell monolayers.
\newblock {\em IEEE Transactions on Nanobioscience}, {\bf 3}(3): 153--163, (2004).


\bibitem{walker04}
D.~C. Walker, J.~Southgate, G.~Hill, A.~Holcombe, D.~R. Hose,
S.~M. Wood, S.~Mac~Neil, and R.~H. Smallwood,
\newblock The epitheliome: agent-based modelling of the social behaviour of cells.
\newblock {\em Biosystems}, {\bf 76}(1-3): 89--100, (2004).

\end{thebibliography}
\end{document}